\documentclass[nofootinbib, aps, twocolumn, superscriptaddress, groupedaddress]{revtex4}  

\usepackage{graphicx}  
\usepackage{dcolumn}   
\usepackage{bm,bbm}        
\usepackage{amssymb}   
\usepackage{amsmath}
\usepackage{amsfonts}
\usepackage{array}
\usepackage{xcolor}
\usepackage{cleveref}
\usepackage{enumerate}

\usepackage{tikz}
\usetikzlibrary{shapes.geometric, arrows}
\usepackage{lipsum}
\usepackage{float}
\usepackage{multirow}
\usepackage{hhline}
\usepackage{tabularx}
\usepackage{subfigure}

\tikzstyle{startstop} = [rectangle, rounded corners, minimum width=3.2cm, minimum height=0.6cm,text centered, draw=black, fill=red!30]
\tikzstyle{io} = [trapezium, trapezium left angle=70, trapezium right angle=110, minimum width=2cm, minimum height=0.6cm, text centered, draw=black, fill=blue!30]
\tikzstyle{process} = [rectangle, minimum width=3cm, minimum height=1cm, text centered, draw=black, fill=orange!30]
\tikzstyle{decision} = [draw, ellipse,fill=green!20, node distance=2cm,
    minimum height=3em]
\tikzstyle{arrow} = [thick,->,>=stealth]

\newcommand{\cO}{\mathcal{O}}
\newcommand{\tD}{\tilde{D}}

\newcommand{\cet}{\chi_{\text{E3}}}
\newcommand{\ch}{\text{ch}}
\newcommand{\Td}{\text{Td}}

\newenvironment{problem}[2]
    {
    \\[4pt]
    
    \begin{tabular}{p{\dimexpr#1}p{\dimexpr0.975\linewidth-#1}}
    \textbf{#2:} &
    }
    {
    \end{tabular}
    \\[4pt]    
    }

\begin{document}

\title{K\" ahler Moduli Stabilization and the Propagation of Decidability}
\author{James Halverson}
\author{Michael Plesser}
\affiliation{Department of Physics, Northeastern University \\ Boston, MA 02115-5000 USA} 
\author{Fabian Ruehle}
\affiliation{CERN Theory Department, 1 Esplanade des Particules, CH-1211 Geneva, Switzerland}
\affiliation{Rudolf Peierls Centre for Theoretical Physics, University of Oxford\\ Parks Road, Oxford OX1 3PU, UK} 
\author{Jiahua Tian}
\affiliation{Department of Physics, Northeastern University \\ Boston, MA 02115-5000 USA}

\date{\today}

\begin{abstract}
Diophantine equations are in general undecidable, yet appear readily in string theory. We demonstrate that numerous classes of Diophantine equations arising in string theory are decidable and propose that decidability may propagate through networks of string vacua due to additional structure in the theory.
Diophantine equations arising in index computations
relevant for D3-instanton corrections to the superpotential exhibit propagation of decidability, with new and existing solutions propagating through networks of geometries related by topological transitions. In the geometries we consider, most divisor classes appear in at least one solution, significantly
improving prospects for K\" ahler moduli stabilization across large
ensembles of string compactifications.
\end{abstract}

\maketitle

\section{Introduction}
Many problems arising in string theory are computationally hard, which could be dynamically
relevant. While a thorough investigation of the question of how hard they are is still in its infancy, a number of results are known. In \cite{Denef:2006ad}, which introduced computational complexity into studies of the string landscape, it was shown that a version of the Bousso--Polchinski model \cite{Bousso:2000xa}
for the cosmological constant is NP-complete. Furthermore, computing the scalar potential in string theory, finding its critical points, and establishing that they are metastable vacua all require solving
NP-hard or co-NP-hard problems~\cite{Halverson:2018cio}. Within the last two years, machine learning was introduced into the study of string theory problems~\cite{He:2017aed,Ruehle:2017mzq,Krefl:2017yox,Carifio:2017bov} to make progress on computationally hard problems.

Even worse, problems arising in physics may be undecidable. In this paper we focus on the Diophantine undecidability, which is the solution to Hilbert's tenth problem. In modern language, the problem is to find an algorithm that solves the following decision problem:
\begin{problem}{0.25\linewidth}{HILBERT}
	Given a polynomial $D(x_1,\dots,x_s)= 0$ with integer coefficients,
	does there exist a solution in the integers?
\end{problem}
$D(x_1,\dots,x_s)= 0$ is a Diophantine Equation. Famously, no such algorithm exists, and therefore HILBERT is undecidable. The proof, which shows that every recursively enumerable set is
Diophantine, is due to Matiyasevich \cite{Matijasevic:1970aaa}, building on work \cite{10.2307/1970289} of Robinson, Davis, and Putnam; the complete result is known as the MRDP theorem.

\medskip

The way in which Diophantine equations arise in string compactifications is by the appearance of integer topological data, such as Chern classes, that describe the compactification space, internal gauge fluxes, and the cycles wrapped by branes. Since this data fixes some physical observables of the string theory compactification, decision problems involving Diophantine equations abound in studies of the string landscape. Even more broadly, by the MRDP theorem \emph{any} decidable or partially decidable decision problem can be turned into a Diophantine equation, and therefore any yes / no question one might ask about the physics of string theory implicitly introduces another Diophantine equation into its study. 

The presence of computationally hard or undecidable problems can affect the dynamics of the underlying system. For instance, folding proteins \cite{Unger1993,bryngelson1994funnels} and relaxing spin glasses \cite{Barahona_1982,doi:10.1142/0271} are both associated with NP-hard problems, which correlates with the existence of proteins and spin glasses that relax on timescales exponential in the system size. Such considerations can give an increased understanding of the dynamics and necessitate additional explanations, including  why our proteins fold quickly, while general proteins do not. For physics examples, the question can be posed of how the Universe ``found'' the state it is currently in if this requires solving a hard or undecidable problem. For instance, when applied to finding a small cosmological constant in an eternally inflating universe \cite{Denef:2006ad,Denef:2017cxt,Bao:2017thx,Halverson:2018cio,Khoury:2019yoo}, this gives a measure that differs significantly from others \cite*{Garriga:2005av,Bousso:2008hz,Freivogel:2011eg,Carifio:2017nyb}.

Given that Diophantine equations arise in string theory in numerous ways and also the size \cite{Bousso:2000xa,Denef2004,Denef:2004cf,Taylor:2015xtz,Halverson:2017ffz,Taylor:2017yqr}
of the string landscape, one might imagine that the resulting equations can be diverse enough to be up against Diophantine undecidability \cite{Cvetic:2010ky}. Of course, this formal concern is notoriously difficult to quantify, since we do not yet have a handle on all of the Diophantine problems of physical interest in string theory.

The focus of this work is to study cases in which Diophantine equations from string theory are easier to decide than the MRDP theorem might naively suggest. That is, given a class of seemingly intractable Diophantine equations that arise for a physical problem in string theory, does the additional structure of the theory render them decidable? We will see this is sometimes the case.

Since there may be transitions between
string vacua / compactifications, for instance due to topology or flux change, it also natural
to consider the landscape as a network where edges represent the transitions, see, e.g., \cite{Carifio:2017nyb}. In this framework,
Diophantine equations on one node may be related to Diophantine equations on another, in which
case it is natural to study if decidability of one implies decidability of the other.
We refer to this as the \emph{propagation of decidability} through the network. 

Though we analyze the decidability of many problems in string theory, our primary
focus is on one important example, for the sake of concreteness.
We study the stabilization of K\" ahler
moduli in string compactifications. These scalar fields are massless at tree-level, but
may be stabilized by non-perturbative corrections to the superpotential, 
specifically worldsheet instantons in heterotic theories or Euclidean D3-instantons (E3)
in the type IIB / F-theory compactifications that we study.
The existence and structure of the superpotential correction depends crucially on the 
spectrum of instanton zero modes, of which there are many, but we focus on a constraint \cite{Witten:1996bn} on
the holomorphic Euler character $\chi(\hat D,\cO_{\hat D})=1$, where $\hat D$ is a divisor in an elliptically
fibered Calabi-Yau fourfold $X\xrightarrow{\pi}B$. This constraint is equivalent \cite{Bianchi:2011qh}
to an index related to a divisor $D\subset B$ where $\hat D = \pi^{-1}(D)$, 
\begin{equation}
\label{eq:chiE3simp}
\chi_{\text{E3}} = -\frac12 \int_B c_1(B)\wedge D \wedge D = 1,
\end{equation}
with Poincar\' e duality implied. Expanding $D$ in an integral basis $D_i$ of the second cohomology $H^{1,1}(X,\mathbbm{Z})$,
\begin{equation}
\label{eq:ExpansionDivisor}
D=\sum_{i=1}^{h^{1,1}} n_i D_i\,,
\end{equation}
it may be expressed as a quadratic Diophantine equation in the integers $n_i$, which is the central object of our study.

\medskip

Our main result is that networks of base geometries $B$ connected by topological transitions
realize the propagation of decidability. Specifically, for a geometry $\tilde B$ that is a blowup of 
another base $B$, solutions to $\cet = 1$ on $B$ significantly aid in determining solutions
to $\tilde \chi_\text{E3}=1$ on $\tilde B$. This applies to the largest known ensembles
of such bases, which exhibit similar physical features and have $2.96\times 10^{755}$ 
and $O(10^{3000})$ geometries \cite{Halverson:2017ffz,Taylor:2017yqr}. Clearly these 
extremely large (but finite) networks are intractable by brute force techniques,
but propagation of decidability nevertheless allows for concrete statements about instanton
solutions. Furthermore, one ensemble \cite{Halverson:2017ffz} may be randomly
sampled from a uniform distribution, which is utilized to show that on average $99.2\%$ percent of 
divisor classes appear as a component in some divisor with $\cet = 1$, even utilizing only
the simplest solutions, suggesting that (up to thoroughly discussed caveats) K\" ahler moduli stabilization
across large ensembles of string compactifications may be easier than naively expected.

Our analytic derivations relating solutions on $\tilde B$ and $B$ likely give rise to many 
more solutions in the networks, but some will depend on details of large sequences of blowups
that introduce model-dependence. We leave a statistical analysis of this type for future work.

\medskip

This paper is organized as follows.
In Section~\ref{sec:physicsimplications} we review known theorems about Diophantine equations,
including decidability of certain cubics and all quadratics. We use them to show that numerous
physical Diophantine problems, often of a geometric nature, are decidable.
In Section~\ref{sec:propagate} we study the E3-index on varieties related by blowup, and
use associated recursion relations that demonstrate the propagation of solutions.
In Section~\ref{sec:KMS} we derive concrete implications of the results of Section~\ref{sec:propagate} for K\" ahler moduli stabilization in type IIB / F-theory compactifications.
In section~\ref{sec:conc}, we conclude.

\section{Physics Implications of Known Diophantine Results \label{sec:physicsimplications}}

In this section we discuss numerous physical applications in string theory of known Diophantine results. 

Physically relevant Diophantine equations in string theory are often of relatively
low degree. This arises because compactification of the extra dimensions in string, M-, or F-theory
often involve a complex algebraic variety $X$ with $\dim_\mathbb{C}(X) \leq n$,
and the Diophantines can arise from intersection theory on $X$.
Such Diophantines are of degree $n$ or less. It is sometimes the case that the intersection of interest on $X$ can be realized within a fixed subvariety $V$ with $\dim_\mathbb{C}(X)\leq d$, in which case the degree is $d$ or below. We will restrict to the case of quadratic and cubic Diophantine equations\footnote{Linear Diophantine equations can be solved in polynomial time~\cite{Lenstra1982:aaa}.}, and then discuss their application in string theory. 

In fact, any set of Diophantine equations can be written as a single quartic Diophantine equation~\cite{Skolem:1938aaa}, albeit in (many) more variables. This is based on the observation that an arbitrary Diophantine equation can be written as a system of quadratic Diophantine equations by introducing auxiliary variables. Then, a set of quadratic Diophantine equations can be turned into a single quartic Diophantine equation by taking the sum of the squares of each individual quadratic equation; since the squares are non-negative, the resulting equation will have a solution iff the original set of quadratic Diophantine equations had a solution. However, given the MRDP result, this makes it very hard to make general statements about quartic Diophantine equations.

\subsection{Quadratic Diophantines in String Theory}

A quadratic Diophantine equation is of the form
\begin{equation}
Q(x_1,\dots,x_s)=0,
\end{equation}
where the polynomial $Q(x_1,\dots, x_s)\in \mathbb{Z}[x_1,\dots,x_s]$
is of degree $2$. The equation may be rewritten as 
\begin{equation}
a_{ij} x_i x_j + h_i x_i = n,
\label{eq:Qcoeff}
\end{equation}
and $H = \text{max}\{|a_{ij}|, |h_i|, n\}$ is known as the height.

Quadratic Diophantine equations are decidable due
to a result of Siegel \cite{MR0311578}. One way
in which this arises is due to the existence of search bounds.
$\Lambda_s(H)$ is a \emph{search bound} if the existence of an
integral solution $(x_1,\dots,x_s)$ to \eqref{eq:Qcoeff} requires
that there is a solution with $|x_i| \leq \Lambda_s(H)$ for $1\leq i \leq s$. Siegel proved that there is a search bound for
any number of variables $s$, and thus quadratic Diophantine
equations are decidable. Though Siegel's search bounds
grow exponentially in $H$, later results  \cite{doi:10.1112/S0024611502013898} demonstrate the existence
of search bounds that are polynomial in $H$ for all $s\geq3$. 
That is, one can always decide existence of a solution to a quadratic Diophantine equation by searching through a finite set of possibilities that grows polynomially (or exponentially for $s=2$) in the maximum of the absolute values of its coefficients. 

Despite being decidable, determining whether there is a solution to a Quadratic Diophantine equation is hard.
For instance, even in the two-variable case  
\begin{equation}
ax_1^2 + b x_2 + c = 0,
\end{equation}
determining whether there is a solution $(x,y)\in \mathbb{Z}^2$ it is NP-complete
via reduction from 3SAT \cite{Manders:1976:NDP:800113.803627}.

A few examples of physically relevant quadratic Diophantines, to which
these results can be applied, are:
\begin{itemize}
\item \textbf{D3-charge.} Seven-branes with gauge bundles
    give rise to induced D3-brane charges that appear in consistency conditions for the theory.

    Specifically, given a type IIB / F-theory
	compactification on a Calabi-Yau threefold $X$ with a
	seven-brane on a divisor $D$ and a target D3-brane charge $T\in \mathbb{Z}$, is there a worldvolume flux
	$L$ (a line bundle) that induces a D3-brane charge $T$?
	This may be studied via an integral parameterization of $L$ 
	as
	\begin{align}
	c_1(L) = \sum_i n_i\, \sigma_{i},
	\end{align}
	where $\sigma_{i}$ is a basis for $H^{1,1}(D,\mathbb{Z})$. Then the answer to the decision problem is yes if and only
	if 
	\begin{equation}
	\int_D \ch_2(L) = T,
	\end{equation}
	where a standard computation expresses the left-handed size
	as a quadratic Diophantine equation in the variables $n_i$.

\item \textbf{Chiral 3-7 modes.} Instanton zero modes crucially affect
	the structure of non-perturbative corrections to the $4d$ $\mathcal{N}=1$ superpotential. Some of these zero-modes are so-called 3-7
	strings \cite{Blumenhagen:2009qh} between a Euclidean D3-instanton and a spacetime filling
	7-brane, and the correction depends crucially on whether there
	are chiral modes of this type \cite{Blumenhagen:2006xt,Ibanez:2006da,Florea:2006si}.

	Specifically, given a type IIB / F-theory
	compactification on a Calabi-Yau threefold $X$ with a
	seven-brane on a divisor $D$, is there a divisor $\tilde D$
	such that a Euclidean D3 brane on $\tilde D$ with instanton flux
	$L$ gives rise to no chiral 3-7 modes at the intersection
	$C = D \cdot \tilde D$? Integrally parameterizing $D$
	as a linear combination $D=m_i D_i$ of effective divisors $D_i$
	and $L$ in
	a way similar to before, answering the question is
	equivalent to solving
	\begin{equation}
	\int_C \Td(C)\, \ch(L|_C \otimes K_C^{1/2}) = 0,
	\end{equation}
	where the left-hand side may be expressed as a quadratic
	Diophantine equation in $n_i, m_i$.

	Related issues are discussed in \cite{Cvetic:2010ky}.
	
\item \textbf{Bianchi Identities.} Given a heterotic string compactification on a Calabi-Yau threefold $X$ with a vector bundle $V$, we need to solve the Bianchi identities for the three-form field $H$, which can be written as
\begin{align}
\ch_2(X)-\ch_2(V)=0\,.
\end{align}
In many cases that have been studied, the bundle $V$ is described as a sum of line bundles, a monad bundle of line bundle sums, or an extension bundle of line bundles. In all cases, we can specify the line bundles via their first Chern classes on the $h^{1,1}(X)$ divisors $D_i$ of $X$, $L_i=\mathcal{O}_X(k_i^1,\ldots, k_i^{h^{1,1}})$. The Bianchi identities then become a (set of coupled) quadratic Diophantine equations in the integers $k_i^a$ (note that decidability as discussed above only applies to a single quadratic Diophantine equation, not to a system of equations).

\item \textbf{GLSM anomalies.} Given a 2d $\mathcal{N}=(0,2)$ gauged linear sigma model, the U(1) charges $q_i^a$ of all defining fields $\Phi_i$ under the $a$ U(1) factors have to be chosen such that the GLSM anomalies vanish. This leads to a (set of coupled) quadratic Diophantine equations in the $q_i^a$ (which can be chosen integral upon changing the U(1) normalization).

\end{itemize}
Of course, many more physical examples could be produced from index formulae in a similar way.  The main
one that we will study in this paper has to do with instanton corrections on divisors $D$ in a K\" ahler
threefold $B$, where in index $\cet = -\frac12 \int_B c_1 \wedge D \wedge D$ defines a quadratic Diophantine
equation in the integers parameterizing $D$.

\subsection{Cubic Diophantine equations in String Theory}

There are also a collection of interesting results for cubic Diophantine equations $C(x_1,\dots,x_s)=0$.
When there is a solution to $C=\partial_i C=0\,\,\, \forall i$, we will say that $C$ is singular; if not, it is non-singular.
When the equation is homogeneous of degree 3, we will say that it is a cubic form.
Results of interest include:
\begin{itemize}
\item A cubic form has a non-trivial solution \cite{10.2307/2414498} if $s\geq 16$.
\item Therefore, cubic forms are decidable if $s\geq 16$. 
\item A non-singular (disallowing the trivial solution) cubic form has a solution \cite{10.1112/plms/s3-47.2.225} if $s\geq 10$. 
\end{itemize}
Given these results about cubic forms and the previous ones about quadratic Diophantine equations, it
is natural to try to make further progress by decomposing a general cubic Diophantine equation as
\begin{equation}
C(x_1,\dots,x_s) = F(x_1,\dots,x_s) + H(x_1,\dots,x_s),
\end{equation}
where $F(x_1,\dots,x_s)$ is a cubic form and $H(x_1,\dots,x_s)$ is a quadratic polynomial.
Then $C(x_1,\dots,x_s)$ is decidable if any one of the following $3$ conditions holds \cite{DavenportH.1964NCE}:
\begin{itemize}
\item $v(F(x_1,\dots,x_s)) \geq 17$.
\item $s\geq 15$ and $4\leq v(F(x_1,\dots,x_s)) \leq s-3$.
\item $s\geq 14$ and we can factor the cubic form as a new variable $\tilde x_n$ times a quadratic form in $n-1$ variables $\tilde x_i$
		\begin{equation}
		C(x_1,\dots,x_s) = \tilde x_n Q(\tilde x_1,\dots, \tilde x_{n-1}),
		\end{equation}
		subject to some additional conditions on the factorization. 
\end{itemize}
These conditions are all, themselves, decidable.

In these conditions, $v(F)$ is an invariant of the cubic form $F$ defined as follows: Given a cubic form, it can be written (non-uniquely) as a sum of products of linear forms $L_i$ and quadratic forms $Q_i$,
\begin{align}
F=\sum_{i=1}^{N} L_i Q_i\,.
\end{align}
Now, $v(F)$ is the minimum number of terms $N$ in this sum for which this decomposition is possible.

\medskip

The simplest results for cubic Diophantine equations are in the case in which they are homogeneous (i.e.\ cubic forms). While such equations sometimes occur in string theory, one often faces more general cubic Diophantine equations, e.g.:

\begin{itemize}
\item \textbf{Existence of Elliptic Fibrations.} It was conjectured by Koll\'ar~\cite{Kollar:2012pv} (which is a proven result by Oguiso~\cite{Oguiso:1993aaa} and Wilson~\cite{Wilson:1994aaa} under some additional assumptions) that a Calabi-Yau threefold $X$ admits an elliptic fibration if $D^3=0$, $D^2\neq0$ and $D\cdot C\geq0$ for all algebraic curves $C\subset X$. Deciding the existence of an elliptic fibration using Koll\'ar's criterion thus requires solving a coupled set of a cubic form, a quadratic form inequality, and a linear condition. 

\item \textbf{Three generations.} In heterotic compactifications on a Calabi-Yau $X$ with a vector bundle $V$ that is given by a sum of line bundles $L_i$, the condition to obtain three net generations of Standard Model Particles is given in terms of the Hirzebruch-Riemann-Roch index theorem and reads
\begin{align}
\label{eq:ThreeNetFamilies}
\begin{split}
\sum_i \chi(L_i,X)&=\sum_i \int_X \Td(TX)\;\ch(L_i)\\
&=\sum_i \frac16 c_1(L_i)^3+\frac{1}{12}c_1(L_i)\;c_2(TX)\\
&\stackrel!=3n\,,
\end{split}
\end{align}
where $n$ is the order of a freely acting symmetry that is commonly used to obtain the Standard Model gauge group. By parameterizing again the line bundles $L_i$ as  $L_i=\mathcal{O}_X(k_i^1,\ldots, k_i^{h^{1,1}})$, this becomes a cubic (inhomogeneous) Diophantine equation. Equation~\eqref{eq:ThreeNetFamilies} computes the net number of quark doublets, but the net numbers of the other particles are given by similar cubic equations.
\end{itemize}

Interestingly, decidability of cubic Diophantines
also arose recently in quantum field theory,
via the derivation \cite{Costa:2019zzy} of the most general solution
to the $U(1)$ anomaly cancellation conditions
in four dimensional $G=U(1)$ theories.

\section{Propagation of Instanton Solutions through Networks of String Geometries \label{sec:propagate}}

We now turn to the central physical problem of this paper: finding Euclidean D3 (E3) instanton corrections to the superpotential across large networks of compactification manifolds in type IIB string theory and F-theory. In cases where a heterotic dual exists, this will also imply the presence of (world-sheet and space-time) instantons in heterotic theories.
Such corrections are of great importance for string cosmology and global dynamics on the landscape.

The detailed structure of an instanton correction depends crucially on its spectrum of zero-modes, and here
we study the simplest case; additional subtleties are discussed in the conclusions. Using various dualities, there are a number of ways to formulate the study of these zero modes. Consider a compactification of M-theory on an elliptically fibered Calabi-Yau fourfold $X \xrightarrow{\pi} B$ with K\" ahler threefold base $B$. In \cite{Witten:1996bn}, Witten
showed that if $X$ is smooth, and M5-brane instanton on a vertical divisor $\hat D$ in $X$ satisfying $h^i(\hat D,\cO_{\hat D})=(1,0,0,0)$ contributes
to the superpotential. Such instanton divisors have holomorphic Euler character 
\begin{align}
\label{eq:holEulerChar}
\chi(\hat D,\cO_{\hat D})=\sum_i (-1)^i\, h^i(\hat D,\cO_{\hat D}) = 1\,.
\end{align}

An F-theory compactification may be obtained by taking M-theory on $X$ in the limit of vanishing fiber, in which case the M5-instanton correction
becomes and E3-instanton correction, arising from an E3 wrapped on $D\subset B$ where $\hat D = \pi^{-1}(D)$. In this duality frame,
 one would like to rewrite the condition $\chi(\hat D,\cO_{\hat D})= 1$ in terms of data intrinsic to $D$ and $B$. This was done via a Leray spectral sequence
 in work of Koll\'{a}r \cite{10.2307/1971390} that relates
 cohomology on $\hat D$ to cohomology on $D$. The result
 matches a detailed IIB instanton zero mode count \cite{Bianchi:2011qh},
 which defined a new index $\chi_{\text{E3}}$ related to cohomology on~$D$. If $B$ is $\mathbbm{P}^1$ fibered, one can moreover apply heterotic / F-theory duality, and $\hat D$ captures world-sheet as well as space-time instanton contributions. In order to not rely on a rational ruling of $B$ in the following, we will focus on E3 instantons, but keep in mind implications for heterotic theories if their dual exists.
 
Witten's instanton condition becomes
\begin{align}\label{eq:chiE3}
	\chi(\hat D,\cO_{\hat D}) = \chi_{\text{E3}} = \chi(D,\mathcal{O}_D) - \chi(D,K_X) = 1.
\end{align}
We choose to utilize $\cet$ written in terms of data on $B$ rather than 
$\chi(\hat D,\cO_{\hat D})$ written in terms of data on $X$. We will see that
this simplifies calculations.

\medskip

In order to address decidability issues, we would like to formulate the condition as a
natural decision problem relevant for determining instanton corrections to the superpotential:
\begin{problem}{0.25\linewidth}{E3-INDEX}
	Given a smooth threefold $B$, is there an effective divisor $D$ with $\chi_{\text{E3}} = 1$?
\end{problem}
Inserting the expansion of $D$ in Equation~\eqref{eq:ExpansionDivisor} into Equation~\eqref{eq:chiE3simp}, the decision problem E3-INDEX becomes a Diophantine equation of degree two in $h^{1,1}(B)$ variables $n_i$. Given $D$ with $\chi_{\text{E3}}=1$, we may say that $D$ or the associated set of $n_i$ provides a yes-solution to E3-INDEX, i.e.\ the $n_i$'s solve the Diophantine equation. 

Since $\cet$ is quadratic, E3-INDEX is decidable, a significant improvement from the general case of Diophantine undecidability. Of course, a decision problem being decidable does not mean that it is tractable. For instance, a smooth toric F-theory base is presented in the appendix, for which the associated $\cet$ is given in \eqref{eqn:nasty}. To the naked eye, this seems intractable, even though it is decidable. To estimate the brute force
tractability, we use the search bounds of \cite{doi:10.1112/S0024611502013898},
\begin{equation}
\Lambda_s(H) = C_4(s) H^{5s + 19 + 74/(s-4)} \,\,\, \text{when}\,\,\, s\geq 5,
\end{equation}
where in the appendix example $s=h^{1,1}=9$ and therefore $\Lambda_{11}(H)\propto H^{78.8}$. Since \eqref{eqn:nasty} has $H=44$, taking $C_4(9)\geq 1$ would give a search set $S_\text{search}$ with
\begin{equation}
|S_\text{search}| \geq (2\times 44^{78.8}+1)^{9} \simeq 10^{1200}.
\end{equation}
Using the concrete quadratic decidability result is already intractable in this $h^{1,1}=9$ case, which is actually quite a low value for $h^{1,1}$. For instance, a generic base in the Tree ensemble \cite{Halverson:2017ffz} has $h^{1,1}\simeq 2000$, in which case brute force search using the search bound becomes even worse. One must find another way to solve E3-INDEX.

\medskip

Our central idea is to utilize additional structure to aid in solving E3-INDEX,
in particular the fact that two 3-dimensional varieties $\tilde B$ and $B$ may be related to one another by topological transitions. Specifically, we will study how the index $\tilde\chi_{\text{E3}}$ of a divisor $\tD$ on $\tilde{B}$ is related to $\chi_{\text{E3}}$ of a divisor $D$ on $B$, where $\tilde{B}$ and $B$ are related via the blowdown $\pi:\tilde{B}\rightarrow B$. The relation is captured by the difference $\Delta\chi = \tilde\chi_{\text{E3}} - \chi_{\text{E3}}$, and pullbacks under $\pi$ will be implicit throughout.

This approach is in the spirit of Mori theory, where a given variety is to be understood from the perspective of simpler models from which it arises via birational transformations. To this end, a related decision problem is
\begin{problem}{0.25\linewidth}{E3-INDEX-PROP}
	Given a blowdown $\pi: \tilde B \to B$ with $\tilde B$ and $B$ smooth, is there an effective divisor $\tilde D$ on $\tilde B$ such that $\tilde\chi_{\text{E3}}=1?$
\end{problem}
Clearly any solution to this problem is a solution to E3-INDEX. However, often the blown down variety is simpler to analyze since $h^{1,1}$ and hence the search space become exponentially smaller, so we would study E3-INDEX-PROP on $\tilde B$
by its relation to E3-INDEX on $B$. Specifically, under which circumstances does a solution of E3-INDEX on $B$ persists to $\tilde B$? Alternatively, when do properties
of $B$ ensure that a new solution exists on $\tilde B$? 

To aid in answering these questions,
 we will study two cases of E3-INDEX-PROP on $\tilde B$:
\begin{itemize}
	\item Case 1: E3-INDEX on $B$ has a solution.
	\item Case 2: The blowup locus $V\subset B$ has $\chi_V = 1$,
\end{itemize}
where $\chi_V:=\chi(V,\cO_V)$ is the holomorphic Euler character of the blowup
locus $V$.

\begin{figure}[t]
	\begin{tikzpicture}[node distance=2cm]
		\node (s1) [startstop] {Is there a $D\subset B$ with $\chi_{\text{E3}}=1$?};
		\node (s4) [io, below of=s1, xshift=-2cm, yshift=0.7cm] {Case 1};
		\node (s6) [decision, below of=s4, xshift=2cm, yshift=-1cm] {E3-INDEX on $\tilde{B}$ has a yes-solution.};
		\node (s8) [startstop, below of=s1, xshift=1.5cm, yshift=0.7cm] {Is $\chi_V = 1$?};
		\node (s5) [io, below of=s8, xshift=0cm, yshift=0.5cm] {Case 2};
		\draw [arrow] (s1) -- node[midway,fill=white] {yes} (s4);
		\draw [arrow] (s1) -- node[midway,fill=white]{no} (s8);
		\draw [arrow] (s4) -- (s6);
		\draw [arrow] (s5) -- (s6);
		\draw [arrow] (s8) --node[midway,fill=white] {yes} (s5);
	\end{tikzpicture}\caption{Flow chart of some cases related to the decision problem E3-INDEX-PROP
	on $\tilde B$, which is a blowup of $B$. If the answers to the posed questions are 
	both no, there may or may not be a solution; further study is needed, and a few
	special cases that yield yes-solutions are discussed below.}
	\label{fig:mainflowchart}
\end{figure}
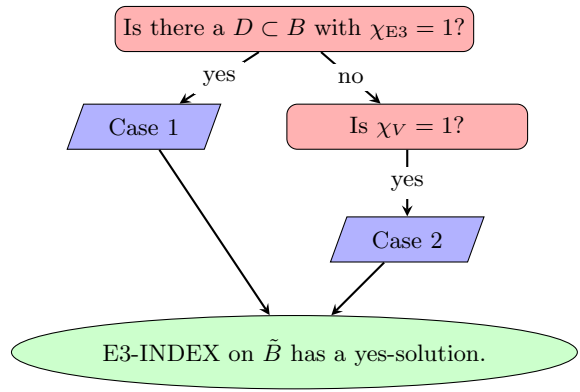
In both cases, we will show that E3-INDEX-PROP on $\tilde B$ has yes-solutions. Case 1 is useful if $B$ is simple enough (i.e.\ the search bounds are low enough) to find solutions to the Diophantine equation~\eqref{eq:chiE3}. Case 2 can be used if E3-INDEX is hard to solve on $B$, but we have additional information about the nature of the blowup. Moreover, the result can be used in some cases to engineer manifolds for which the answer to E3-INDEX is yes, by choosing appropriate blowup loci. We summarize the procedure and propagation of decidability of E3-INDEX in Figure~\ref{fig:mainflowchart}.

In both cases, we will show that E3-INDEX-PROP on $\tilde B$ has yes-solutions, which will motivate the definition of a third decision problem which is decidable, covers most instances of E3-INDEX and yields yes-solutions: 
\begin{problem}{0.25\linewidth}{SEQ-E3-INDEX}
	Given $\tilde B$ that has a sequence of blowdowns $\tilde B\rightarrow \dots \to \hat B \xrightarrow{\pi} B$ such that the blowup locus $V\subset X$ has $\chi_V=1$, is there an effective divisor $\tilde D \subset \tilde X$ with $\tilde\chi_{\text{E3}}=1$?
\end{problem}

\bigskip

Our results and calculations depend critically on whether the blowup is along a curve $C = D_1\cdot D_2$ or at a single point $P = D_1\cdot D_2\cdot D_3$. In the following, we expand a divisor in the blown up variety $\tilde{B}$ as 
\begin{align}
\tD = \bar{D} + n_eE\,,
\end{align} 
where $E$ is the exceptional divisor of the blowup and the divisor $\bar{D}$ can be expanded, analogously to Equation~\eqref{eq:ExpansionDivisor}, as
\begin{align}
\label{eq:ExpansionDbar}
\bar{D} = \sum_{i=1}^{h^{11}}\bar{n}_i\mathcal{D}_i\,.
\end{align}
Here, $\mathcal{D}_i$ is the proper transformation of the divisors $D_i$. When it causes no confusion we will use the same notation $D_i$ for both the divisor class on $X$ and its pull-back under the blowup.

\smallskip

For a blowup at $P$ we have:
\begin{align}\label{eq:chiE3_point}
	\tilde\chi_{\text{E3}} = -\frac{1}{2}\bar{D}^2c_1 + D_1D_2D_3n_e^2.
\end{align}
If there already exists a divisor $D$ on $B$ that solves $\chi_{\text{E3}} = 1$, we set $\bar{D} = D$ in Eq.~\eqref{eq:chiE3_point} and see that the only solution to $\Delta\chi = 0$ is $n_e = 0$; the solution $D$ to E3-INDEX on $B$ propagates to a solution of E3-INDEX on $\tilde B$.

Alternatively, if $\bar D^2 c_1 =0$ and the blowup locus is a single point $D_1D_2D_3=1$,  then $\bar D \pm E$ is a solution. For $\bar D=0$ this matches a result in \cite{Witten:1996bn}, but we note that the solution is much more general. For instance, if $\bar D$ is either a K3 surface with trivial normal bundle or if it is a toric divisor associated with a facet interior of a reflexive polytope, the condition $\bar D^2 c_1 = 0$ is satisfied and $\bar D \pm E$ are solutions; others likely exist.

\smallskip

For a blowup along $C=D_1\cdot D_2$, we have
\begin{align}\label{eq:chiE3_curve}
\begin{split}
	\tilde\chi_{\text{E3}} &= -\frac{1}{2}\bar{D}^2c_1 - n_e \, \bar D \cdot C + \chi_C n_e^2 \\
	&= \cet - n_e\, \bar D \cdot C + \chi_C n_e^2
\end{split}
\end{align}
where $\chi_C = \frac{1}{2}D_1D_2(c_1-D_1-D_2)$ is the holomorphic Euler characteristic of the curve $C$. 

Again, if we know a $D$ on $B$ that satisfies $\chi_{\text{E3}} = 1$, we can also use it after the blowup, i.e.\ set $\bar D=D$. The condition $\tilde \chi_{\text{E3}}=1$ is then equivalent to
\begin{align}
	\Delta\chi =\tilde \chi_{E3}- \cet = n_e\, (\chi_C\, n_e - D\cdot C)=0.
\end{align}
Therefore, we see that a solution (besides the pullback of $D$, i.e.\ the $n_e=0$ case) is 
$\tD = D + kE$ where $k = \frac{D\cdot C}{\chi_C}$, where $k$ must be an integer. 

Alternatively, if we set all $n_i=0$ in~\eqref{eq:ExpansionDbar}, the condition~\eqref{eq:chiE3_curve} becomes
\begin{align}
	\tilde\chi_{\text{E3}} = \chi_C n_e^2 = 1,
\end{align}
which is solved only for $\chi_C=1$ and $n_e = \pm 1$.  However,
since $-E$ is not effective, only $E$ is a solution. That is,
when $B$ is blown up along a genus $0$ curve, $E$ is a solution to E3-INDEX
on $\tilde B$. 

Finally, if $\bar{D}^2 c_1 = 0$ but $\bar D \neq 0$,
\begin{equation}
\cet = n_e\, (\chi_C n_e - \bar D \cdot C) = 1,
\end{equation}
which requires
\begin{equation}
n_e = \pm 1\,, \qquad \qquad \chi_C = 1-g(C)= 1\pm \bar D\cdot C\,.
\end{equation}
This can be solved many ways for different choices of $g(C)$ and $\bar D$, e.g.
\begin{itemize}
\item If $g(C)=1$, then $\bar D \pm E$ is a solution if $\bar D \cdot C = \mp 1$. 
\item If $g(C)=0$, then $\bar D \pm E$ is a solution provided $\bar D \cdot C = 0$. 
\end{itemize}
The latter solution can be non-trivial due to self-cancellation in $\bar D \cdot C$.  For instance, if $c_1\cdot C =0$
then $(D_1+D_2)\cdot C = -2$ since $g(C)=0$ and $\bar D = D_1 + D_2 + D_3 + D_4$
makes $\bar D \pm E$ a solution provided $(D_3+D_4)\cdot C = 2$. This sounds contrived, 
but it arises naturally in toric geometry contexts: if $C$ is a toric curve meeting
at the intersection of two three-cones with associated divisors $(D_3,D_1,D_2)$ and $(D_4,D_1,D_2)$
and the four divisors $D_{1,2,3,4}$ are facet interiors, then all of the conditions are met.

To summarize, some guaranteed solutions to E3-INDEX-PROP on $\tilde B$ are:
\begin{enumerate}[(S1)]
	\item The pullback of a solution from $B$ to $\tilde B$. \label{S1}
	\item The exceptional divisor $E$. It is always a solution in the point blowup case,
	and also in the curve blowup case if $g(C)=0$. \label{S2}
	\item $\bar D+kE$ in the curve blowup case if it is effective and $\bar D$ is a solution
	on $B$, where $k=\bar D \cdot C/\chi_C$.\label{S3}
	\item $\bar D\pm E$, where $\bar D$ is not necessarily a solution on $B$, if
	certain conditions hold. In the point blowup case, the condition is that $\bar D^2 c_1=0$.
	In the curve blowup case, $\bar D \pm E$ is a solution provided that $\bar D^2 c_1=0$ and  $\chi_C = 1-g(C) = 1\pm \bar D \cdot C.$ \label{S4}
\end{enumerate}
We have labeled the classes (S1)-(S4) to allow for simplified discussions. Solutions of class (S1) and (S2) are fairly automatic and correspond to Case $1$ and Case $2$ above, whereas those of class (S3) and (S4) requires 
some additional non-trivial checks. There could be other
interesting classes of guaranteed solutions, as well.

\medskip

We wish to also take into account the fact that string geometries arise in
large networks. The transition from $B$ to $\tilde B$ we discussed, along with its associated
solutions, are only two nodes and one edge in that network. One would like to know
how solutions propagate through the entire network, rather than just from one node to
its neighbor. 

This question is addressed by the decision problem SEQ-E3-INDEX defined above. By the results of Case $1$ and Case $2$, SEQ-E3-INDEX is not only decidable, but has yes-solutions. Case $2$ guarantees the existence of a solution to E3-INDEX on $\hat X$ when blowing up at a curve $C$ with $\chi_C = 1$ or when blowing up at a point, and then case $1$ applied to $\hat X$ guarantees a solution on $\tilde X$. That is, SEQ-E3-INDEX is always decidable, and yes-solutions always exist.

\subsection{Concrete solutions to SEQ-E3-INDEX\label{sec:concreteEE}}
We showed above that the exceptional divisor $E$ of a blowup along $V\subset X$ is always a solution to E3-INDEX if $\chi_V = 1$. We now want to see whether there exists a solution of the form $E_1 + kE_2$ after a second blowup where $E_1$ is the exceptional divisor of the first blowup and $E_2$ is that of the second blowup. 

There four possible combinations, since either of the blowups could be the blowup of a point or of a curve. However, if the second blowup is a point blowup, we must have $k=0$. Hence,  we consider the two cases where the second blowup is along a curve. 

\medskip
\noindent\textbf{First blowup is at a curve.} We first consider cases where the first and the second blowup are all blowups along a curve. There are three such cases. The first blowup is along a curve $C = D_1\cdot D_2$. For the exceptional divisor $E_1$ of the first blowup to solve E3-INDEX on $\tilde X$ we require $\chi_C = 1$.

\smallskip
\textbf{Blowup 1:} The first case is the sequence of blowups:
\begin{align}
	\tilde X \xrightarrow{(E_2|E_1,D_3)} \hat X\xrightarrow{(E_1|D_1,D_2)} X,
\end{align}
where the notation $(E_1|D_1,D_2)$ denotes a blowup along $C=D_1\cdot D_2$ with
exceptional divisor $E_1$, and similarly for $(E_2|E_1,D_3)$.
We have
\begin{align}
	\tilde \chi_{\text{E3}} = 1 + D_1D_2D_3(k^2+k).
\end{align}
Since $\chi_C = 1$, we see that $k=0$ is a solution to $\tilde \chi_{\text{E3}} = 1$, as is $k=-1$ provided that $D_1D_2D_3=1$, which is always the case in toric examples where these
three divisors form a three-cone. Therefore we see that besides the pullback of $E_1$ on $\tilde X$, $E_1 - E_2$ is a solution to E3-INDEX under mild assumptions.

\smallskip
\textbf{Blowup 2:}  The second case is the sequence of blowups:
\begin{align}
	\tilde X \xrightarrow{(E_2|E_1,\tilde D_1)} \hat X\xrightarrow{(E_1|D_1,D_2)} X.
\end{align}
We have:
\begin{align}
	\tilde \chi_{\text{E3}} = 1 - (D_2 \cdot C) \,k + k^2.
\end{align}
We see that the solutions to $\tilde \chi_{\text{E3}} = 1$ are 
\begin{align}
	k=0 \qquad\text{or}\qquad k = D_2 \cdot C\,.
\end{align}
Therefore, we see that besides the pullback of $E_1$ on $\tilde X$, $E_1 + (D_2 \cdot C) E_2$ is also a solution to E3-INDEX, provided that it is an effective divisor.
It is not uncommon that $D_2\cdot C = -1$, for instance, if $D_2$ is itself the exceptional divisor of a blowup along a curve that meets $D_1$ at a point.

\smallskip
\textbf{Blowup 3:} The third case is the sequence of blowups:
\begin{align}
	\tilde X \xrightarrow{(E_2|\tilde D_1,D_3)} \hat X\xrightarrow{(E_1|D_1,D_2)} X.
\end{align}
We have:
\begin{align}
	\tilde \chi_{\text{E3}} = 1 - D_1D_2D_3k + \chi_{C'}k^2
\end{align}
where $C' = D_1\cdot D_3$; we emphasize that $C'$ is a curve on the original space that is not a blowup locus. We see that the solutions to $\tilde \chi_{\text{E3}} = 1$ are 
\begin{align}
	k=0 \qquad \text{or} \qquad k = \frac{D_1D_2D_3}{\chi_{C'}}\,.
\end{align}
Therefore, we see that besides the pullback of $E_1$ on $\tilde X$, $E_1 + \frac{D_1D_2D_3}{\chi_{C'}} E_2$ is also a solution to E3-INDEX. 

\medskip
\noindent\textbf{First blowup is at a point.} There are two cases where the first blowup is at a point and the second is along a curve. In both cases, for the first exceptional divisor to be
a solution, the first blowup is at a single point, given by $D_1D_2D_3 = 1$.

\smallskip
\textbf{Blowup 1:} The first case is the sequence of blowups:
\begin{align}
	\tilde X \xrightarrow{(E_2|\tilde D_1,\tilde D_2)} \hat X\xrightarrow{(E_1|D_1,D_2,D_3)} X.
\end{align}
We have:
\begin{align}
	\tilde \chi_{\text{E3}} = 1-k + \chi_C k^2
\end{align}
where $C = D_1\cdot D_2$; we emphasize that $C$ is a curve on the original space that is not a blowup locus. We see that the solutions to $\tilde \chi_{\text{E3}} = 1$ are 
\begin{align}
	k=0 \qquad \text{or} \qquad k = \frac{1}{\chi_C}\,.
\end{align}
Therefore, we see that besides the pullback of $E_1$ on $\tilde X$, $E_1 + \frac{1}{\chi_C} E_2$ is also a solution to E3-INDEX, provided that it is effective, which requires $\chi_C=1$.

\smallskip
\textbf{Blowup 2:} The second case is the sequence of blowups:
\begin{align}
	\tilde X \xrightarrow{(E_2|\tilde D_1,E_1)} \hat X\xrightarrow{(E_1|D_1,D_2,D_3)} X.
\end{align}
We have
\begin{align}
	\tilde \chi_{\text{E3}} = 1+k+k^2.
\end{align}
 We see that the solutions to $\tilde \chi_{\text{E3}} = 1$ are $k = 0,-1$. Therefore we see that besides the pullback of $E_1$ on $\tilde X$, $E_1 - E_2$ is also a solution to E3-INDEX.

\section{Implications of propagation for K\"ahler moduli stabilization \label{sec:KMS}}

Having formulated various decision problem related to instanton corrections to the
superpotential and having found different classes of solutions, we wish to discuss
the implications for the stabilization of K\" ahler moduli.

This is subtle, because as we have emphasized, the presence of additional zero modes
not captured by the condition $\cet=1$ can alter the structure of the correction. However,
given the importance of K\" ahler moduli stabilization for
realistic string cosmology, it behooves us to proceed with a discussion under the assumption
that additional zero modes do not kill the corrections associated with our solutions. This
will allow us to discuss how many K\" ahler moduli appear in the superpotential.
A detailed study of the assumption, and also the importance of instantons that do not have 
$\cet=1$, such as in \cite{Bianchi:2011qh}, are interesting directions for future work.

For the purposes of K\" ahler moduli stabilization, a more
relevant question is ``How many K\" ahler moduli appear in at
least one instanton correction?". This can be encoded in a counting problem
\begin{problem}{0.25\linewidth}{NAIVE-STABLE}
	Given threefold $B$ and all divisors $D\subset B$ with $\cet=1$, how many K\"ahler moduli
	appear in at least one of the instanton corrections?
\end{problem}
Equivalently, what is the rank of the lattice spanned by the divisors? This is the ``naive" stability count because of the above assumption, and also because a single correction does not guarantee that the modulus is stabilized, for instance due to  a runaway. However, since it can be studied across large networks of geometries, it is a good starting point.

\medskip

For simplicity, let us study the concrete case of networks of toric varieties
that serve as bases of F-theory elliptic fibrations. Consider the so-called Tree ensemble \cite{Halverson:2017ffz},
which forms a single-component connected network
of $2.96 \times 10^{755}$ toric threefolds. The network is formed by
recursively performing blowups of a fixed initial
weak-Fano toric threefold $B$ along toric curves,
which are $\mathbb{P}^1$'s, and toric points. Either
of these blowup loci $V$ have $\chi_V=1$, and therefore
any geometry $\tilde B$ in the network (provided
it is not the initial threefold $B$) has a sequence
of blowdowns $\tilde B\to \dots \to \hat B \to B$.
That is, all but one of the $10^{755}$ geometries in the Tree ensemble
provide instances of SEQ-E3-INDEX, and therefore have at least one instanton solution.

For K\" ahler moduli stabilization, one would rather like to know the answer to
NAIVE-STABLE on $\tilde B$. Since the number of K\" ahler moduli $h^{1,1}(B)$ on the variety $B$ and $h^{1,1}(\tilde B)$ on the variety $\tilde B$ are fixed,  and any
single blowup increases $h^{1,1}$ by 1, all directed paths in the network between $B$ and $\tilde B$
are necessarily of the same length. However, the toric blowups only have loci $V$ with
$\chi_V=1$, and therefore the exceptional divisor associated to each blowup in the sequence
yields a solution to $\tilde \chi_\text{E3}=1$ on $\tilde B$,
i.e.\ at least 
\begin{equation}
 \Delta h^{1,1}(\tilde B):=h^{1,1}(\tilde B) - h^{1,1}(B)
\end{equation}
 K\" ahler moduli satisfy NAIVE-STABLE. The corrections are schematically of the form
 \begin{equation}
 	W = \sum_{i=1}^{h^{11}(\tilde B)} A_i(\phi) \, e^{-2\pi \, T_{E_i}} + \dots
 \end{equation}
where 
\begin{align}
T_{E_i} = \int_{E_i} \frac12 J\wedge J + i C_4 
\end{align}
is the complexified
K\" ahler modulus associated to the exceptional divisor $E_i$, $J$ is the K\" ahler form,
and $C_4$ is the Ramond-Ramond four-form.

The probability that a K\"ahler modulus on $\tilde B$ satisfies
NAIVE-STABLE is therefore at least $\Delta h^{11}(\tilde B)/h^{11}(\tilde B)$.
In the Tree ensemble, the expected percentage of K\" ahler
moduli naively stabilized using just exceptional divisors is
\begin{equation}
\label{eqn:percentnaivestable}
100\%\times \mathbb{E}_{\tilde B\sim U_\text{tree}}\left[\frac{\Delta h^{11}(\tilde B)}{h^{11}(\tilde B)}\right] = 100\%\times \frac{2448}{2483}=98.6\%,
\end{equation}
where $U_\text{tree}$ is the uniform distribution on the
Tree ensemble. This could be improved even further by the inclusion of 
divisors pulled back from $B$ in mixing terms, as we will discuss momentarily.

We emphasize that all of these results also apply to the so-called Skeleton
ensemble \cite{Taylor:2017yqr}, which is less
restrictive than the tree ensemble and is estimated
to have $O(10^{3000})$ elements.
Generally, these results apply to \emph{any} network of toric varieties constructed
by recursively performing toric blowups from an initial
toric $B$, which are automatically along loci with $\chi_V=1$.

\bigskip
Let us discuss be more details by studying both mixing terms and potential caveats.

One caveat to note is that if $E_i$ intersects a later blowup locus in the sequence, the
nature of the blowup could render a generic representative of class $E_i$ a reducible variety,
for instance with normal crossing singularities. In such a case a careful zero mode analysis
is required to determine whether
\begin{enumerate}
\item there are additional zero modes present at the normal crossing locus and
\item additional physics gives interactions to lift these modes.
\end{enumerate}
This has not been studied in detail in the literature.

However, if an exceptional divisor $E$ is involved in $N$ additional blowups with
exceptional divisors $E_i$, the divisor $E-\sum_{i=1}^N E_i$ can be irreducible and does not suffer from normal crossing singularities.
To that end, one would like to understand under which conditions such divisors solve $\tilde \chi_\text{E3}=1$.
For simplicity suppose that $N=1$, and consider $\tilde B \xrightarrow{\pi} B$ with exceptional divisor
$\tilde E$ and another divisor $E$ that is the exceptional divisor of a previous blowup along
a locus $V$ with $\chi_V=1$. The latter condition guarantees that $E$ solves $\cet=1$ on
$B$, and therefore it also solves $\tilde \chi_\text{E3}=1$ on $\tilde B$. Since $E$
is a solution, having a new solution of the form $E + k\tilde E$ with $k\neq 0$ requires
that the blowup locus of $\pi$ be a curve $\tilde{C}\subset B$, cf.\ (S3). A solution of the proposed form
requires
\begin{equation}
k = \frac{E\cdot \tilde{C}}{\chi_{\tilde{C}}}=-1,
\end{equation}
where for the sake of irreducibility we took $k=-1$.
In the toric case, where $g(\tilde{C})=0$, this amounts to
requiring that $E\cdot \tilde{C}=-1$.

Corrections of the type that we just discussed induce \emph{mixing terms}, which
could give rise to important competition (such as in racetrack scenarios) and couplings between K\" ahler moduli in the
scalar potential. Any given blowup such as $\tilde B \xrightarrow{\pi} B$ in a sequence of blowups
allows to study solutions of type (S1)-(S4) for that transition, i.e.\ whether the exceptional
divisor $E$ of the blowup gives rise to a superpotential mixing of that K\" ahler modulus
with some of the others. The pullback solutions (S1) and the exceptional divisor solution (S2)
do not, but by their very form, solutions of type (S3) and (S4) give rise to mixing. A detailed
study of the prevalence of (S3)- and (S4)-type solutions could give crucial information on
K\" ahler modulus mixing, but is beyond the scope of this paper. 

We note, however, that in the previous section we described toric cases in which such solutions exist. For instance, consider a toric divisor $D$ on the original base $B$ in the Tree
ensemble, which is associated with a fine regular star triangulation of a reflexive polytope. These are not contributing to the simple estimate~\eqref{eqn:percentnaivestable}, and we would like to improve the situation. On the biggest 3d polytopes, which dominate the ensemble,
$16$ of the $38$ toric divisors correspond to facet interiors, and therefore satisfy $D\, c_1 = 0$. While these do not satisfy $\cet=1$ on $B$, after any blowup involving
$D$, $\tilde D= D-E$ is a solution. Nearly all geometries in the Tree ensemble have
such a blowup, and therefore the moduli corresponding to facet interiors 
raise the expectation for NAIVE-STABLE, 
\begin{equation}
\label{eqn:percentnaivestable2}
100\%\times \mathbb{E}_{\tilde B\sim U_\text{tree}}\left[\frac{\Delta h^{11}(\tilde B)}{h^{11}(\tilde B)}\right] = 100\%\times \frac{2464}{2483}=99.2\%,
\end{equation}
where the moduli that we have still not taken into account correspond to edge interiors and vertices of the polytope.

Mixing terms are also important from another perspective: they might allow for a correction
involving a K\" ahler modulus $T$ that does not appear on its own in any instanton correction,
which is therefore crucial for the stabilization of $T$. Consider, for instance, the case where
$B$ is the result of a blowup along a curve $C$ with $g(C)\neq 0$. Then by (S2), the exceptional
divisor $E$ is \emph{not} a solution to $\cet =1$ on $B$. Working on $B$, one could try
searching for solutions of type (S3) or (S4) which potentially give rise to
an instanton contribution involving $E$. Alternatively, if one passes to  $\tilde B \xrightarrow{\pi} B$ with exceptional divisor $\tilde E$ and blowup locus $\tilde C$, then $E-\tilde E$ solves
$\tilde \chi_\text{E3}=1$ on $\tilde B$ provided that
\begin{equation}
g(\tilde C) = 1-g(C) + E\cdot \tilde C
\end{equation}
and $E-\tilde E$ effective. If one wants $\tilde E$ to automatically
be a solution, $g(\tilde C)=0$ and the condition simplifies to
\begin{equation}
E\cdot \tilde C = 1-g(C),
\end{equation}
which is a search problem for an appropriate $\tilde C$.

\section{Discussion and Summary of Results \label{sec:conc}}

We studied Diophantine equations that arise in string theory with
regard to their decidability and physics.

In Section \ref{sec:physicsimplications}, we reviewed known theorems about
Diophantine equations. The central theorems for our applications are that 
all quadratic Diophantines are decidable, as are cubic form Diophantines in more
than $15$ variables. We presented
simple examples in which quadratic and cubic Diophantine equations appear in string theory.
In the quadratic case, we presented examples related to D3-brane charge,
chiral 3-7 instanton zero modes, Bianchi identities, and GLSM anomalies. In the cubic case,
examples included Koll\'ar's condition for the existence of an elliptic fibration in
a Calabi-Yau threefold, as well as obtaining three generations of Standard Model chiral
multiplets in heterotic line bundle compactifications.

\medskip
In Section \ref{sec:propagate}, we focused on a Diophantine equation which
is relevant for instanton corrections to the superpotential. More specifically, for an
elliptic Calabi-Yau fourfold $X\xrightarrow{\pi}B$, M5-brane instanton corrections may arise
from divisors $\hat D$ in $X$ satisfying Equation~\eqref{eq:holEulerChar}. In this case, the instanton
should be thought of in the type IIB / F-theory limit, where it is a Euclidean D3-brane instanton
wrapped on $D$. In cases where a heterotic dual exists, the corresponding instantons are world-sheet and space-time instantons. 
Since the instanton condition leads to a quadratic Diophantine equation, the question whether there exists a suitable $D$ is decidable. We call this decision problem E3-INDEX.

Our central idea is to make progress on E3-INDEX by considering how the problem behaves
under topological transition. That is, given a geometry $\tilde B$ that is the blowup of a geometry
$B$, may E3-INDEX on $\tilde B$ be understood in terms of the problem on $B$? From this perspective,
one thinks of a network of string geometries related by topological transitions, which are represented
by edges, and asks whether decidability of E3-INDEX on one node (geometry) decides E3-INDEX on
an adjacent node. This is the propagation of decidability.

We formalize it in terms of a decision problem E3-INDEX-PROP, which introduces the blowup into the problem
and the notion that solutions on $B$ may be related to those on $\tilde B$, based on the structure
of the topology change. We find two rather useful subcases of the problem. In Case 1,
E3-INDEX on $B$ itself has a solution $D$. For both point and curve blowups, that solution pulls back
to a solution on $\tilde B$, i.e.\ it has $\tilde \chi_\text{E3}=1$ on $\tilde B$. Additionally, in the case of a blowup along a curve $C$, an
extended solution for E3-INDEX on $\tilde B$ exists under mild assumption. It is of the form
$D + k\, E$ where $k=(D\cdot C)/\chi_C$ must be an integer and the divisor must be effective. This is
often true in toric varieties. In Case $2$, we simply note that the exceptional divisor itself is
a solution to E3-INDEX on $\tilde B$ if $\chi_V:= \chi(V,\cO_V)=1$, where $V$ is the blowup locus in
$B$. This is the case if $V$ is a genus $0$ curve or a point.

The results of the two cases motivate a third decision problem, SEQ-E3-INDEX, which postulates
that the variety of interest $\tilde B$ may be obtained by a sequence of blowups from a blowup of an
initial variety with blowup locus satisfying $\chi_V=1$. Then by the results of Case 1 and Case 2,
all instances of SEQ-E3-INDEX have a yes-solution, i.e. a divisor with $\tilde \chi_\text{E3}=1$ on $\tilde B$.

We derive numerous classes of solutions to E3-INDEX-PROP on $\tilde B$, sometimes with mild assumptions. We categorize them according to whether
they are pullbacks of solutions from
$B$ (called (S1)), the exceptional divisor of the blowup (S2), of the form $D+kE$ where $D$ is a solution on $B$ (S3), or 
of the form $D' \pm E$ where $D'$ is not necessarily a solution on $B$.

In subsection \ref{sec:concreteEE} we derive solutions that depend only on two exceptional divisors in a sequence of blowups,
for instance $E-\tilde E$, with $E$ the exceptional divisor of the first blowup. If $E$ itself is a solution, then a necessary
condition for $E+n_{\tilde e} \tilde{E}$ to be a solution is that the second blowup is a curve blowup, so that interesting solutions
involving both $E$ and $\tilde E$ arise in the cases of point-then-curve blowup sequences, or curve-then-curve blowup sequences.
There are five total possibilities, which we study in detail and present solutions accordingly.

\medskip

In section \ref{sec:KMS} we discuss the implications for K\" ahler moduli stabilization, which are of great importance for string cosmology and
global dynamics on the landscape.

We define a problem called NAIVE-STABLE that counts the number of K\" ahler moduli
that appear in some instanton correction. Our results immediately
demonstrate that all exceptional divisors in well-studied ensembles of toric
geometries generated by blowups are necessarily solutions, which make giving 
lower bounds on NAIVE-STABLE straightforward. Some mixing terms, associated with
facet interiors, are also easy to take into account.

By studying NAIVE-STABLE, we find that $99.2\%$ of K\" ahler moduli
appear in some instanton correction, on average, across the Tree ensemble.
Additional possibilities for mixing terms are also studied.

\medskip
This provides ample motivation for future work. 

On the formal side, there is additional research that must be done into aspects
of instanton physics that are not fully understood. One has to do with instantons wrapped on divisors
with normal crossing singularities, which appear in our work and \cite{Braun:2017nhi}. Open questions include whether the singularities give rise to zero modes
and whether additional physics may lift the zero modes. There is also the question of
a general formalism for understanding the role of 3-7 modes in instanton corrections,
though progress has been made in specific
contexts, e.g., \cite{Blumenhagen:2010ja,Cvetic:2011gp,Cvetic:2012ts,Kerstan:2012cy,Martucci:2015oaa}. Lifting deformation modes, for instance
via instanton fluxes \cite{Bianchi:2011qh,Grimm:2011dj}, could also be crucial in stabilizing moduli that do not arise in any divisor with $\cet=1$.

On the statistical / data side,
it would be interesting to systematically explore the toric ensembles with respect to the instanton
solutions that we have derived. Our analytic analysis and estimates are only the
simplest possible cases, and there is likely much more structure to be understood.

\medskip

More broadly, we have demonstrated a case in which structure in string theory
allows for the avoidance of worst-case complexity. In the present case, we
showed that geometric structures in string theory allows to decide Diophantine equations that arise in string theory, despite them being undecidable in general. It would be interesting to find other cases where physical problems in string theory are easier than might naively be expected, both for practical reasons and questions concerning the dynamics of string vacua.

\bigskip
\noindent \textbf{Acknowledgments.}
We thank Paddy Fox, Justin Khoury, Cody Long, Muyang Liu, Benjamin Sung, and Yi-Nan Wang for discussions.  J.H. is supported by NSF CAREER grant PHY-1848089.

\clearpage
\appendix

\section{An example of propagation}

We construct a smooth toric variety $X_{P_8}$ that is the result of 8 blowups on $\mathbb{P}^3$. The fan associated with $X_{P_8}$ has the following rays:
\begin{align*}
\begin{array}{l@{~}l@{~~}l@{~~}}
	v_0 = (1,0,0),\ &v_1 = (0,1,0),\ &v_2 = (0,0,1), \\
	v_3 = (-1,-1,-1),\ &e_1 = (1,1,1),\ &e_2 = (1,1,2), \\
	e_3 = (1,1,3),\ &e_4 = (1,1,4),\ &e_5 = (1,0,1), \\
	e_6 = (1,0,2),\ &e_7 = (1,0,3),\ &e_8 = (1,0,4).
\end{array}
\end{align*}
We will use the same notation for both the rays and their corresponding coordinates. From the linear relation between the rays, we see that $X_{P_8}$ is obtained from $X_{P_0} = \mathbb{P}^3$ via the following sequence of blowups:
\begin{align}
	\begin{aligned}
		X_{P_8}&\xrightarrow{(e_8|v_2,e_7)}X_{P_7}\xrightarrow{(e_7|v_2,e_6)}X_{P_6}\xrightarrow{(e_6|v_2,e_5)}X_{P_5} \\
		&\xrightarrow{(e_5|v_0,v_2)}X_{P_4}\xrightarrow{(e_4|v_2,e_3)}X_{P_3}\xrightarrow{(e_3|v_2,e_2)}X_{P_2} \\
		&\xrightarrow{(e_2|v_2,e_1)}X_{P_1}\xrightarrow{(e_1|v_1,v_2,v_3)}X_{P_0}=\mathbb{P}^3
	\end{aligned}
\end{align}
We denote the 9 generators of the K\" ahler cone (and by abuse of notation, the Poincar\' e
dual divisors) by $G_0, G_1, \dots, G_8$. A divisor $D\subset X_{P_8}$ can be expanded as $D = \sum_{i=0}^8n_iG_i$. 

Using the results of Section \ref{sec:propagate}, we know that the exceptional divisor $E_8$ corresponding to the last blowup 
\begin{align}
X_{P_8}\xrightarrow{(e_8|v_2,e_7)}X_{P_7} 
\end{align}
is a solution to $\chi_{\text{E3}} = 1$ on $X_{P_8}$. We can express $E_8$ as a linear combination of $G_i$'s:
\begin{align}\label{eq:relation_E8}
	E_8 = -G_0 + G_8\,.
\end{align}

We also compute $\chi_{\text{E3}}$ on $X_{P_8}$, which is a quadratic polynomial in the 9 $n_i$'s. We have:
\begin{align}
	\chi_{\text{E3}} =&\ -18 n_0^2-22 n_1 n_0-33 n_2 n_0-44 n_3 n_0-42 n_4 n_0\nonumber \\
	&\ -12 n_5 n_0-11 n_6 n_0-40 n_7 n_0-38 n_8 n_0-5 n_1^2\nonumber \\
	&\ -12 n_2^2-22 n_3^2-21n_4^2-2 n_5^2-n_6^2-20 n_7^2-19 n_8^2\nonumber \\
	&\ -16 n_1 n_2-22 n_1 n_3-33 n_2 n_3-22 n_1 n_4-33 n_2 n_4\nonumber \\
	&\ -44 n_3 n_4-8 n_1 n_5-12 n_2n_5-16 n_3 n_5-15 n_4 n_5\nonumber \\
	&\ -5 n_1 n_6-8 n_2 n_6-11 n_3 n_6-11 n_4 n_6-4 n_5 n_6\nonumber \\
	&\ -22 n_1 n_7-33 n_2 n_7-44 n_3 n_7-42 n_4n_7-14 n_5 n_7\nonumber \\
	&\ -11 n_6 n_7-22 n_1 n_8-33 n_2 n_8-44 n_3 n_8-42 n_4 n_8\nonumber \\
	&\ -13 n_5 n_8-11 n_6 n_8-40 n_7 n_8\,.
\label{eqn:nasty}
\end{align}

Using Equation~\eqref{eq:relation_E8} it is easy to check that following set of integers solves $\chi_{\text{E3}} = 1$:
\begin{align}
	&-n_0 = n_8 = 1\ \text{and}\ n_i = 0,\ \forall i\neq 0,8\,.
\end{align}
This agrees with our result that $E_8$ solves $\chi_{\text{E3}} = 1$ on $X_{P_8}$ via the analytic algebro-geometric method.

\bibliography{refs}

\end{document}